\documentclass[%
 reprint,
groupedaddress,
showpacs,preprintnumbers,
 amsmath,amssymb,
 aps,
 prl,
twocolumn,
]{revtex4-1}

\usepackage{graphicx}% Include figure files
\usepackage{dcolumn}% Align table columns on decimal point
\usepackage{bm}% bold math
\usepackage{amsmath}
\usepackage[colorlinks=true]{hyperref}
\newcommand{\mean}[1]{\mbox{$\langle{#1}\rangle$}}

\begin{document}

%\linenumbers

%\preprint{APS/123-QED}

\title{An Adiabatic Phase-Matching Accelerator}% Force line breaks with \\
%\thanks{This work was supported by the European Research Council (ERC) under the European Union's Seventh Framework Programme 
%(FP/2007-2013)/ERC Grant agreement nunder the European Union's Seventh Framework Programme 
%(FP/2007-2013)/ERC Grant agreement no. 609920}

\author{F. Lemery$~^{1}$, K. Floettmann~$^1$,  P. Piot~$^{3,4}$, F. X. K\"{a}rtner $^{1,2}$, R. A{\ss}mann$^{1}$ \\
$^1$ DESY, Notkestrasse 85, 22607 Hamburg, Germany\\
$^2$ Department of Physics, University of Hamburg, Jungiusstra{\ss}e 9, 20355 Hamburg, Germany\\
$^3$ Department of Physics, and Northern Illinois Center for Accelerator \& Detector Development, \\
Northern Illinois University DeKalb, IL 60115, USA \\
$^4$ Fermi National Accelerator Laboratory, Batavia, IL 60510, USA}

\begin{abstract}

We present a general concept to accelerate non-relativistic charged particles. Our concept employs an adiabatically-tapered dielectric-lined waveguide which supports accelerating phase 
velocities for synchronous acceleration. We propose an ansatz for the transient field equations, show it satisfies Maxwell's equations under an adiabatic approximation and find excellent agreement 
with a finite-difference time-domain computer simulation.  The fields were implemented into the particle-tracking program {\sc astra} and we present beam dynamics results
for an accelerating field with a 1-mm-wavelength and peak electric field of 100~MV/m. The numerical simulations indicate that a $\sim 200$-keV electron beam can be accelerated to an energy of 
$\sim10$~MeV over $\sim 10$~cm. The novel scheme is also found to form electron beams with parameters of interest to a wide range of applications including, e.g., future advanced accelerators, and 
ultra-fast electron diffraction.

\end{abstract}

\pacs{29.27.-a, 41.85.-p, 41.75.Fr}% PACS, the Physics and Astronomy

\date{\today}
\maketitle
High-energy charged-particle accelerators have emerged as invaluable tools to conduct fundamanetal scientific research.
Circular high energy colliders continue exploring nuclear and high-energy landscapes, searching for hints beyond the standard model.  
Linear accelerators capable of forming high-quality electron bunches have paved the way to bright, coherent X-ray sources to probe ultrafast 
phenomena at the nanometer-scale with femtosecond resolutions in condensed matter, life science and chemistry. Accelerators have also found 
medical applications such as, e.g., high-resolution imaging and oncology.

Modern klystron-powered conventional accelerators incorporate radio-frequency (RF) accelerating structures optimized to provide suitable accelerating fields 
typically in the frequency range $f \in[0.1, 10]$~GHz (i.e. wavelengths respectively in the range $\lambda \in[3, 0.03]$~m). Unfortunately, power requirements and 
mechanical breakdowns in accelerating cavities have limited the permissible electric fields to $E_0\lesssim 50$~MV/m, 
leading to km-scale infrastructures for high-energy accelerators.  These limitations have motivated the development of advanced acceleration 
techniques capable of supporting high accelerating fields. Accelerating structures based on dielectric waveguides or plasmas operating in a
higher-frequency regime [${\cal O}$(THz)] have been extensively explored in the relativistic regime.

A key challenge in accelerating low-energy non-relativistic beams with higher frequencies stems from the difference between the beam's velocity 
and accelerating-mode's phase velocity. This difference leads to ``phase slippage" between the beam and the accelerating field which ultimately limits 
the final beam energy and quality. Scaling to higher frequencies (i.e. shorter wavelengths) 
exacerbates the problem~\cite{kwangje,rosenswag,piot,klausrf}.  A figure of merit conventionally used to characterize the beam dynamics in the 
longitudinal degree of freedom during acceleration of a non-relativistic beam is the normalized vector potential $ \alpha = (e E_0 \lambda)/(2\pi mc^2)$,  
where $e$ and $mc^2$ are respectively the electronic charge and rest mass, and $E_0$ is the time averaged accelerating field. 
Conventional electron photoinjectors typically operate in a relativistic regime of $\alpha\gtrsim1$; retaining relativistic field strengths while scaling to smaller wavelengths 
(following $E_0\propto\lambda^{-1}$) is challenging beyond RF frequencies but is now routinely attained using high-power infrared 
lasers in plasmas operating at $f \sim 1$~THz. Additionally,  low-$\alpha$ acceleration in e.g. dielectric-laser accelerators (DLA) 
at optical wavelengths is interesting nonetheless owing to the foreseen compact footprints, relatively large gradients and high-repetition rates. 
DLA has pursued side-coupled grating accelerators ~\cite{plettner, breuer} to address phase-slippage with tapered grating periods. Likewise, proton 
accelerators have established radio frequency quadrupoles (RFQs)~\cite{rfq} and drift-tube linacs with comparably long wavelengths to achieve low-$\alpha$ acceleration.

In this Letter, we show analytically that a longitudinally-tapered dielectric-lined waveguide (DLW) can support electromagnetic fields with a 
longitudinally-dependent phase velocity. Therefore, by properly tailoring the spatial taper profile of the DLW, one can establish an electromagnetic 
field  with instantaneous phase velocity $v_p(z)$ matching the beam velocity $\beta(z) c$ along the direction of motion $\hat{z}$~\cite{lemeryIPAC1,lemeryIPAC2}. 
The concept is shown to be able to accelerate a non-relativistic electron beam ($\sim200$~keV) generated out of a compact low-power RF gun to relativistic energies 
$\sim 10$~MeV within a few centimeters. We hypothesize an ansatz for the transient field equations supported in a tapered DLW, show they verify Maxwell's equations 
and validate them against a finite-difference time-domain (FDTD) electromagnetic simulation. We finally implement the transient field equations in the beam dynamics 
program {\sc astra}~\cite{astra}, present start-to-end simulations and validate the concept to form bright electron bunches suited for the production of attosecond 
X rays via inverse Compton scattering~\cite{axsis}. We especially find that a single derived tapered waveguide can have a versatile range of operation,
yielding electron bunches with a broad set of properties of interest for various applications.\\

A cylindrical-symmetric DLW consists of a hollow-core dielectric waveguide (with relative dielectric permittivity $\epsilon_r$) with its outer surface 
metallized~\cite{Frankel1947}. Introducing the  cylindrical-coordinate system $(r, \phi, z)$, where $r$ is referenced w.r.t. the DLW axis along the 
$\hat{z}$ direction, and denoting the inner and outer radii as respectively $a$, and $b$, the electromagnetic field ($\pmb E, \pmb B)$ associated to 
the accelerating (TM$_{01}$) has the following non-vanishing components:
\begin{equation}
\label{eq:EHusual}
\begin{split}
E_z&=E_0 I_0(r k_1)\sin(\omega t -  k_z z + \psi),\\
E_r&=\frac{E_0 k_z}{k_1} I_1(r k_1)\cos(\omega t - k_z z + \psi),\\
B_\phi&=\frac{\omega \epsilon_0 \mu_0 E_0}{k_1} I_1(r k_1)\cos(\omega t - k_z z + \psi),
\end{split}
\end{equation}
where $k_1 \equiv  \omega \sqrt{\frac{1}{v_p^2}-\frac{1}{c^2}}$, $k_2  \equiv \omega \sqrt{\frac{\epsilon_r}{c^2}-\frac{1}{v_p^2}}$, $k_z = \frac{\omega}{v_p}$, 
$I_m(...)$ are the modified $m$-th order Bessel's function of the first kind, $E_0$ is the peak axial field amplitude, $v_p$ is the phase velocity, 
$\omega \equiv 2\pi f $ and $\psi$ is a phase constant.  In the limit $v_p \to c$, i.e. $\lim_{k_1 \to 0} I_1(k_1 r)/k_1 = r/2$ and $\lim_{k_1 \to 0} I_0(k_1 r)=1$,
thus the transverse fields become completely linear and the longitudinal field becomes independent of the transverse coordinate.
Conversely smaller values of $v_p$ result in increasingly nonlinear transverse fields and a strong dependence of $E_z$ on the transverse coordinate. 
This is a general feature of phase velocity matched modes and not restricted to the specific case discussed here.  For high frequency structures
the effect of nonlinearities is however exacerbated because the ratio of typical transverse beam dimensions to the wavelength is larger than in
conventional rf structures.

Solutions of the characteristic equation~\cite{Frankel1947}
yield the allowed $(\omega, k_z)$ for propagating modes and depend on the DLW structure parameters $(a,b,\epsilon_r)$. A propagating mode must have 
a real-valued longitudinal component for the wavevector $k_z$. Correspondingly, $k_2$ sets a limit on the phase velocity of a propagating mode via 
$k_2 < \frac{\omega n}{c}$, or $v_p > \frac{c}{n}$ (where $n\equiv \sqrt{\epsilon_r}$ is the dielectric's index of refraction).  Finally, we note 
that the field amplitudes reduce with decreasing phase velocities ($v_p \to c/n$).

We now turn to modify Eq.~\ref{eq:EHusual} to describe the fields associated to a tapered DLW. Specifically, we hypothesize that the 
transverse dimensions $(a,b)$ at a longitudinal coordinate $z$ locally determine $v_p$ and $E_0(z)$.  In addition, the phase at a position $z$ 
should depend on the integrated phase velocity upstream of the structure. Given these conjectures, we make the following ansatz for the 
non-vanishing ($\pmb E, \pmb B)$ fields
\begin{equation}\label{eq:ansatz}
\begin{split}
E_z&=E_0(z) I_0(r k_1(z))\sin(\omega t - \int_0^z dz k_z(z)  + \psi)\\
E_r&=\frac{E_0(z) k_z(z)}{k_1(z)} I_1(r k_1(z))\cos(\omega t - \int_0^z dz k_z(z) + \psi)\\
B_\phi&=\frac{\omega \epsilon_0 \mu_0 E_0(z)}{k_1(z)} I_1(r k_1(z))\cos(\omega t - \int_0^z dz k_z(z) + \psi),
\end{split}
\end{equation}
where now $k_z(z)$ is integrated from the structure entrance ($z=0$) to the longitudinal coordinate $z$. 
The latter set of equations also introduce an explicit $z$ dependence for  $E_0(z)$, $k_1(z)$, $k_z(z)$. For convenience 
we  define $\Psi(t,z)\equiv \omega t - \int_0^z dz k_z(z) + \psi$.

In order to validate  our ansatz, we check that it satisfies Maxwell's equations, starting with the Amp\`ere-Maxwell law 
$\frac{\partial {\pmb E}}{\partial t} = -{\pmb \nabla} \times {\pmb B}$ which yields
\begin{equation}\label{eq:MA}
\frac{1}{c^2}\frac{\partial E_z }{\partial t}=-\frac{1}{r}\frac{\partial}{\partial r}(rB_\phi). 
\end{equation}
Computing the rhs and lhs of the equation given the field components listed in Eq.~\ref{eq:ansatz} and making use of 
the identity $\frac{\partial}{\partial r} r I_1(k_1 r) = k_1 r I_0(k_1 r)$ confirms that Eq.~\ref{eq:MA} is fulfilled as both 
sides equal $\frac{\omega}{c^2}E_0 I_0(k_1 r) \cos(\Psi(t,z))$. 

Next, we consider Gauss' law ${\pmb \nabla} \cdot {\pmb E}=0$ which yields, for the fields proposed in Eq.~\ref{eq:ansatz}, 
\begin{eqnarray}\label{eq:Gauss}
\frac{\partial}{\partial z}E_z=-\frac{1}{r}\frac{\partial}{\partial r}(rE_r). 
\end{eqnarray}
The rhs of the latter equation gives 
\begin{equation}
\begin{split}\label{eq:eqnLim}
\frac{\partial}{\partial z}E_z &= E_0[-k_z I_0 \cos(\Psi(t,z)) + r k_1' I_1 \sin(\Psi(t,z))]\\ 
 &+ E_0'I_0 \sin(\Psi(t,z))\\
\end{split}
\end{equation}
while its lhs results in 
\begin{equation}\label{eq:eqnLim2}
\begin{split}
-\frac{1}{r}\frac{\partial}{\partial r}(rE_r) &=-\frac{E_0 k_z \cos(\Psi(t,z))}{k_1 r}\frac{\partial}{\partial r}(rI_1(k_1 r))\\
&=-k_z E_0 I_0(k_1 r) \cos(\Psi(t,z)).
\end{split}
\end{equation}
Equations~\ref{eq:eqnLim} and \ref{eq:eqnLim2} are generally not equal unless the tapering of the DLW is sufficiently slow or adiabatic,
\begin{equation}\label{eq:limits}
\begin{split}
\left|\frac{r k_1' I_1(k_1 r)}{k_z I_0(k_1 r)}\right| \ll 1 , \mbox{~and }
\left|\frac{E_0'}{E_0 k_z}\right| \ll 1.
\end{split}
\end{equation}
Equations~\ref{eq:limits} are independent of $\epsilon_r$ and present general conditions to the evolution of a traveling mode in a tapered waveguide which
may also be of interest to plasma acceleration with e.g. smaller field gradients and potentially higher repetition rates.

Insofar, our discussion has solely concentrated on the electromagnetic aspect of the problem. Let us now consider the beam dynamics 
of a charged particle accelerating in a tapered DLW. The requirement for continuous synchronous acceleration imposes $v_p(z)=\beta(z) c$. The 
longitudinal phase space dynamics is described by the coupled ordinary differential equations,
\begin{equation}
\begin{split}\label{eq:diffEq}
\frac{\partial z}{\partial t} &= \beta c,\\
\frac{\partial \beta}{\partial t} &= \frac{e E_0(z)}{\gamma^3 m c} I_0(k_1 r) \sin(\Psi).
\end{split}
\end{equation}
In addition to synchronous acceleration, the transverse-dynamics plays a crucial role in the formation of bright electron beams.
The transverse force can be calculated from the Lorentz force, 
\begin{equation}\label{eq:focusForces}
\begin{split}
F_r&=e(E_r-\beta c B_\phi)\\
&=e E_0 \left(\frac{1}{\beta_p}-\beta\right) k_0 \frac{I_1(k_1 r)}{k_1}\cos(\Psi),
\end{split}
\end{equation}
where $\beta_p\equiv v_p/c$ is the normalized phase velocity.  For synchronous acceleration ($\beta=\beta_p$), the latter equation simplifies to
\begin{equation}
F_r=e E_0 \frac{k_0}{\gamma^2 \beta}\frac{I_1(k_1 r)}{k_1}\cos(\Psi), 
\end{equation}
implying transverse defocusing forces for $\Psi\in[-90,0]$~deg (i.e. the compression phase where the bunch tail experiences a stronger longitudinal field than the head)
and transverse focusing forces for $\Psi\in[-180,-90]$~deg and no transverse force on-crest at $\Psi=-90$~deg. 
We note that the amplitude of the force is increased by a combination of the particle and the phase velocity (Eq.~\ref{eq:focusForces}), while the
nonlinearity of the field is a result of the matching to the phase velocity alone. For $v_p \sim c$, as e.g. in conventional rf guns, (Eq.~\ref{eq:focusForces}) reduces to $F_r=eE_0k_0/(\gamma^2(1+\beta))r/2$;
the transverse fields are linear in $r$ and the longitudinal field is independent of the radial coordinate. In the matched case however, strong and nonlinear fields appear at low energies, which in
combination with the r-dependence of $E_z$, strongly affect the beam dynamics.
The transverse emittance is an important figure of merit which characterizes the phase-space density of the beam defined as 
$\varepsilon_r\simeq \varepsilon_{u}=\frac{1}{mc}[\mean{u^2}\mean{p_u^2}-\mean{u p_u}]^{1/2}$ where $\mean{...}$ is the 
statistical averaging over the beam distribution.
Therefore at injection, the transverse beam size $\sigma_r$ should be minimized to mitigate emittance and energy spread dilutions.

\begin{figure}[t]
\centering
\includegraphics[width=0.49\textwidth]{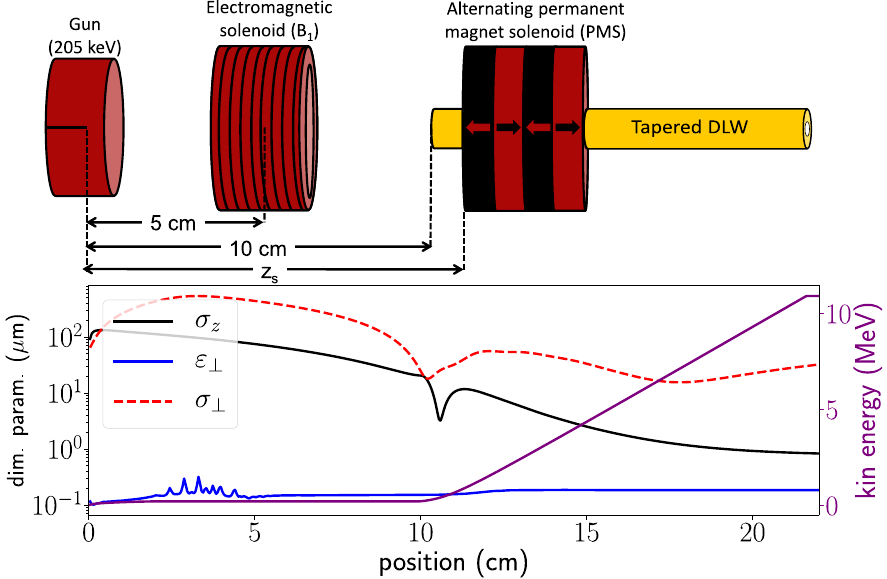}
\caption{
Diagram of the accelerator concept (top) and corresponding evolution of the bunch's transverse emittance ($\varepsilon_r$), rms transverse beam size ($\sigma_r$), longitudinal
bunch length ($\sigma_z$) (all left axis) and the kinetic energy (right axis) along the accelerator beamline (bottom).  The example corresponds to an operating point
($\phi$, $E_0$)=(79.3~deg, 106.875~MV/m); see text for details.}
\label{fig:setup}
\end{figure}

We now describe a start-to-end simulation considering a driving field with $\lambda=1$~mm ($f=300$~GHz) and $E_0=$100~MV/m corresponding to $\alpha\simeq 0.03$. 
A C++ program was developed to integrate the equations of motion (Eqs.~\ref{eq:diffEq}) for one electron given the set of initial conditions: electron injection energy, 
wavelength, peak accelerating field, and DLW geomtery. In parallel, the characteristic equation is solved to derive the appropriate taper.
Consequently, an electron injected on crest will not experience any phase-slippage through the structure.  
Additionally, scaling the accelerating field $E_0 \to \eta  E_0$ offsets the point of zero phase-slippage by an amount $\delta \Psi=\arcsin(1/\eta)$.
We specialize our study to the case where the DLW has a constant inner radius $a$ and devise the outer radius $b(z)$ to ensure synchronous acceleration throughout the DLW. 
The DLW is taken to be made of quartz ($\epsilon_r=4.41$) with length $L=11.5$~cm, and $a=0.5$~mm. This choice of material limits the phase velocity to values $v_p>c/n\sim.48c$
thereby requiring an injection energy $>70$~keV. The group velocity of $\sim$ 0.5~c and length of the structure dictate a required pulse length of $\sim$ 383~ps corresponding
to a pulse energy of 7.2~mJ for $E_0$=100~MV/m.   

We consider a compact, low-power, field-enhanced S-band ($f=3$~GHz) gun~\cite{klausCompactGun} with a photocathode as our electron source.
The gun is operated off-crest to generate short 
$\sigma_z\sim20$~$\mu$m, $205$~keV electron bunches at the DLW entrance with a total charge $Q=100$~fC.  An electromagnetic solenoid 
with variable peak axial magnetic field $B_1$ is located 5~cm downstream of the cathode and focuses the bunch into the DLW structure 
positioned 10~cm from the cathode. To control the strong defocusing forces (Eq.~\ref{eq:focusForces}) during the early stages of acceleration, 
a 4-slab, $\sim$4~cm alternating-permanent-magnet solenoid~\cite{gehrke} with a maximum axial field $B_2$=1.5~T surrounds part of the DLW and is 
located at a distance $z_s$ from the cathode; see Fig.~\ref{fig:setup}(top).  This setup was not globally optimized.
\begin{figure*}[t]
\centering
\includegraphics[width=\textwidth]{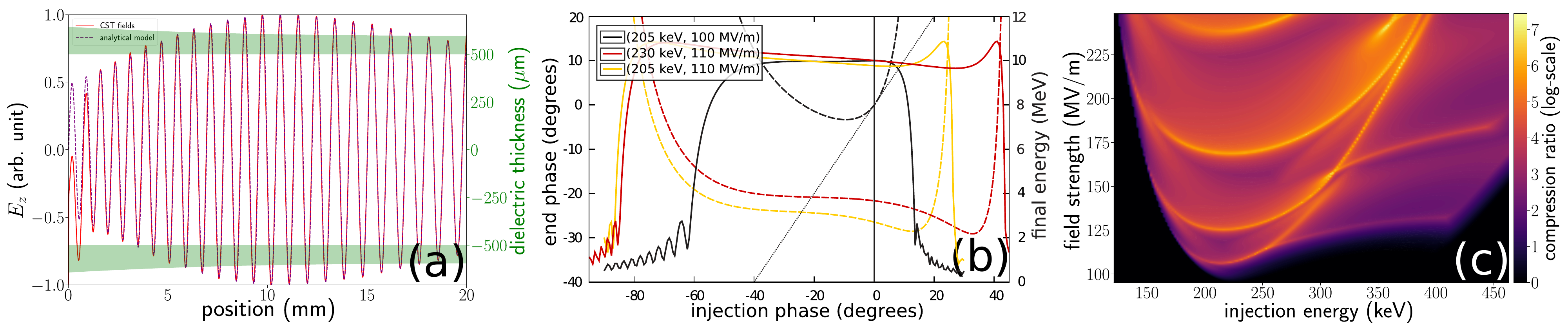}
\caption{(a) Geometry of the dielectric-layer tapering (green shaded area, right axis) over the entrance of the structure; the initial dielectric
thickness is 143~$\mu$m ($v_p=0.7c$) and asymptotically approaches 91~$\mu$m ($v_p=c$).  In addition we show the
comparison between our analytic field ansatz with FDTD code {\sc CST-MWS} over the first 20~mm. (b) Final energy (solid traces, left column) and 
end phase (dashed lines, right column) as a function of 
injection phase for various accelerating gradients and initial kinetic energies. The black diagonal dashed line
shows $\phi_e$=$\phi_i$, intersections with the phase portraits indicate zero phase-slippage. (c) The compression ratio between the injection phase and end
phase, $\Delta \phi_i/\Delta \phi_e$ in log-scale as a function of injection energy and accelerating gradient for an input bunch length spanning 60~deg ($\Delta \phi_i$=60~deg.)}
\label{fig:singlePart}
\end{figure*}
At the entrance of the structure the matched phase velocity is $v_p=0.7c$ and the accelerating gradient is reduced to $E_0\sim20$~MV/m; the adiabatic condition from Eq.~\ref{eq:limits}
for $r$=100~$\mu$m at $z$=0 gives 0.0017 and approaches $10^{-7}$ toward the end of the structure.  For completeness we use the FDTD program 
{\sc CST MWS}~\cite{cst} to simulate the field propagation for the first 2~cm of the DLW where the majority of the taper occurs; here we simulated
a Gaussian THz pulse with 1\% bandwidth, while our simulation in {\sc astra} utilizes a flat-top pulse. The simulated 
fields are in excellent agreement with our semi-analytical field; see Fig.~\ref{fig:singlePart}(a).  Some discrepancies arise at the entrance and 
exit of the structure due to transient effects not included in our model.

We can gain some significant insight into the longitudinal dynamics with a single electron; we illustrate the energy gain and end phase as a function of 
initial phase in Fig.~\ref{fig:singlePart}(b).  Generally, larger accelerating gradients and injection energies than the matched conditions increase the longitudinal acceptance of the structure.
Plateaus in the end phases suggests bunch compression across the flat injection phase width. In Fig.~\ref{fig:singlePart}(c) we show the resulting compression 
ratio in log-scale, $\frac{\Delta \phi_i}{\Delta \phi_e}$, as a function of initial energy $\mathcal{E}_i$ and field strength $E_z$ for an 
input bunch spanning 60 degrees. The phase trajectory through the structure is determined by $\mathcal{E}_i$, $E_z$ and $\phi_i$; changing 
these parameters will alter the phase trajectory and impact the forces experienced by the bunch along the structure; this implies that a 
single structure has a very broad range of operational capabilities.
\begin{figure}[b]
\centering
\includegraphics[width=0.5\textwidth]{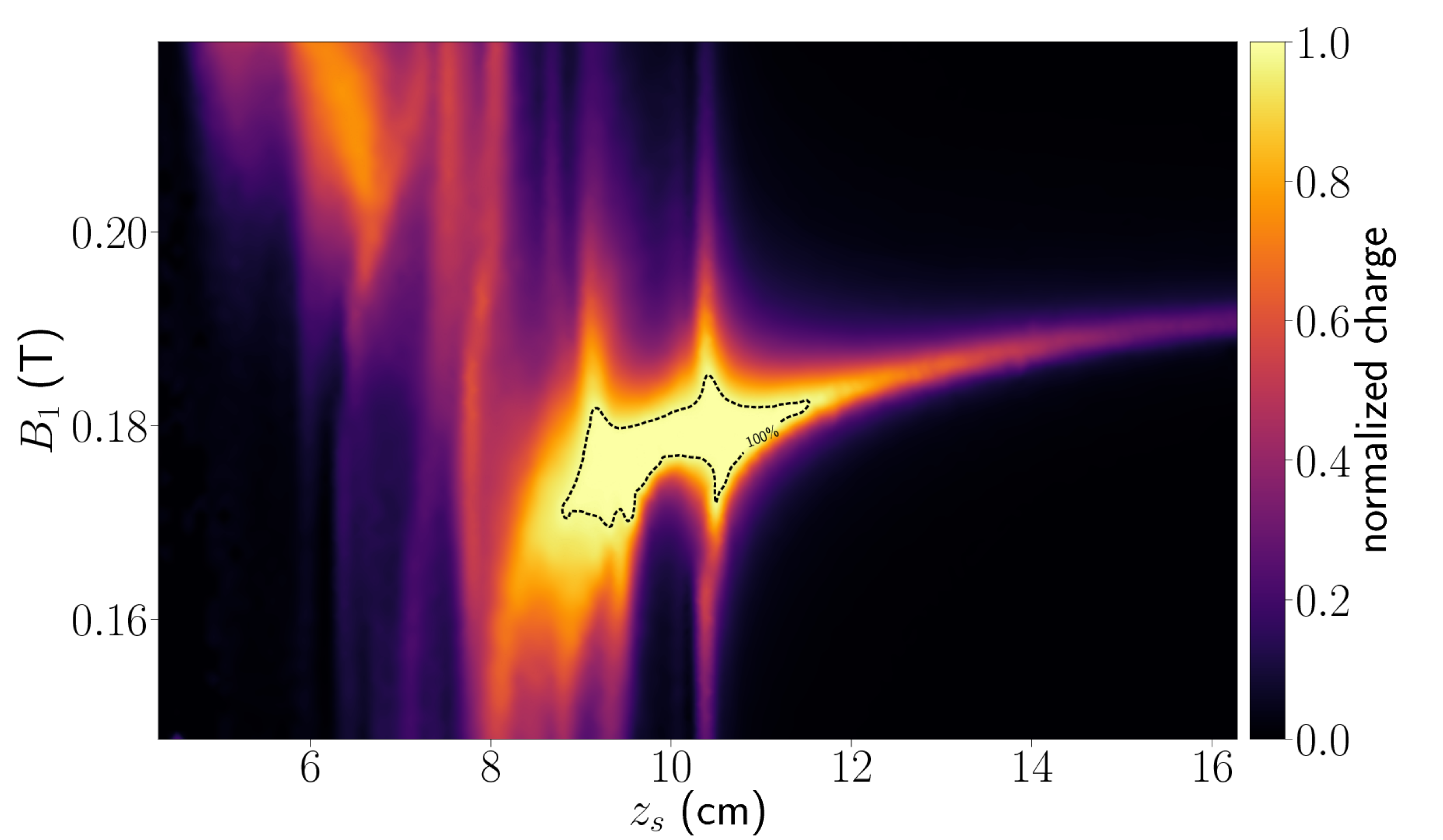}
\caption{Charge fractional transmission through the structure as a function $B_1$ and $z_s$ for injection parameters ($\mathcal{E}_i$, $E_z$)=(205~keV, 105.8~MV/m) 
corresponding to a maximum bunch compression point from Fig.~\ref{fig:singlePart}(c).  A black dashed line encompasses 100\% transmission.}
\label{fig:tranMatch}
\end{figure}

The transverse matching into and through the structure essentially depends on the balance between the transverse defocusing 
forces from the DLW  and focusing optics from the solenoids. Different phase trajectories will generally have different 
transverse forces along the structure. We illustrate the transmission through the structure as a function of our matching 
optics $B_1$ and $z_s$ in Fig.~\ref{fig:tranMatch} associated to a local compression maximum $(\mathcal{E}_i, E_z)$=(205~keV, 105.8~MV/m).  
To accommodate such an injection energy, we accordingly minimize the bunch length by choosing an appropriate field strength and injection 
phase in the gun, $E_{gun}$=113.55~MV/m, $\phi_{gun}$=217 deg. Larger acceptances allow less stringent requirements on the beam matching 
and allows a larger operational range for the same matching point. 

\begin{figure}[ht!]
\centering
\includegraphics[width=0.48\textwidth]{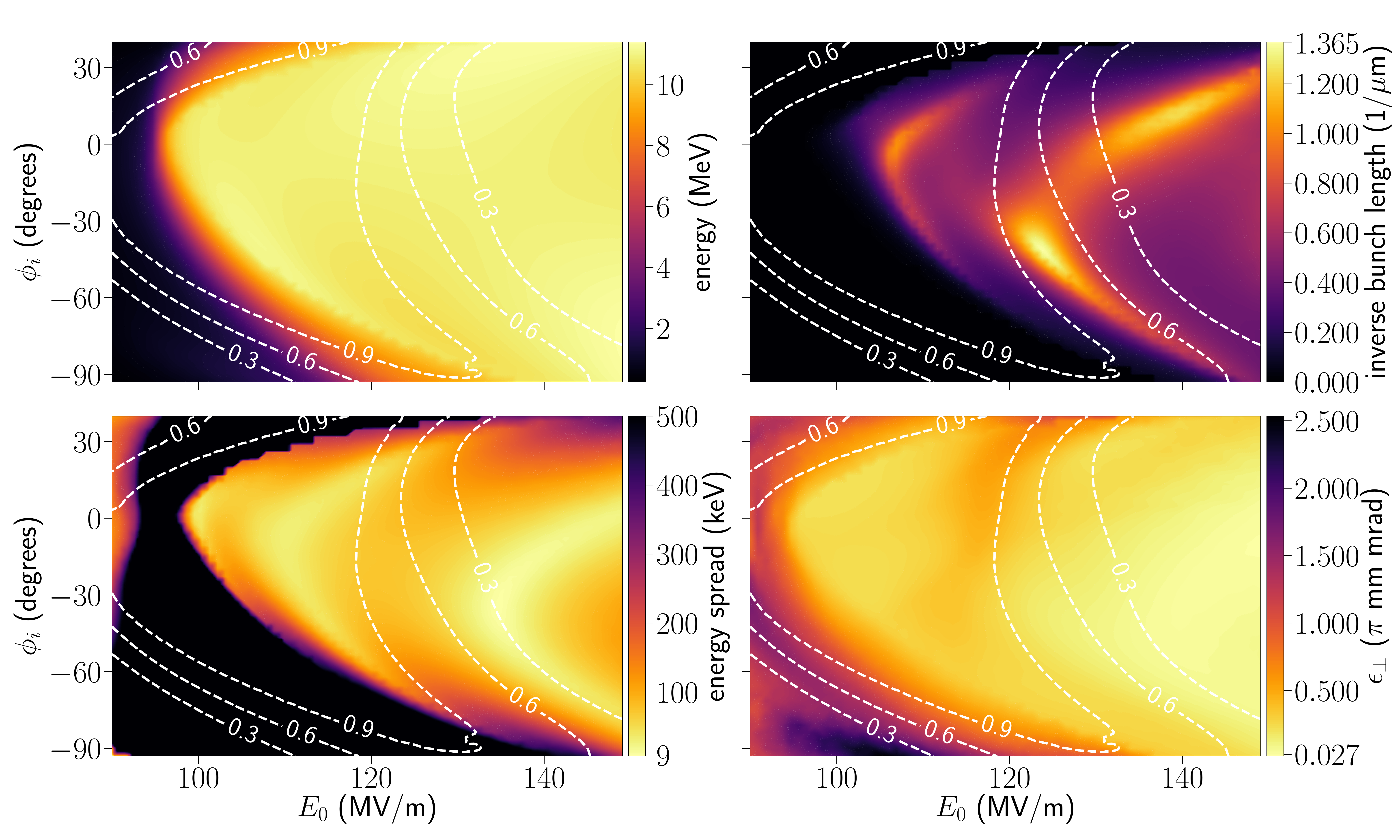}
\caption{Final bunch energy, inverse bunch length, energy spread, and normalized transverse emittance for the matched case, ($B_1$, $z_s$)=(0.179~T, 10.5~cm).
In each case we overlay the final bunch charge as white contour levels for 0.3, 0.6 and 0.9 charge transmission.  While all final energies are approximately equal ($\sim$11~MeV),
the structure allows for the production of widely-tunable electron beams.}
\label{fig:accelDyn}
\end{figure}

Finally, we explore the beam dynamics of the structure for a matching point in Fig.~\ref{fig:tranMatch}, ($B_1$, $z_s$)=(0.179~T, 10.5~cm) and show
the resulting final energy, energy spread, inverse bunch length, and transverse emittance for scans over ($\phi$, $E_0$) in Fig.~\ref{fig:accelDyn}.  In each
figure we include white contour lines representing the final bunch charge; in cases with large offsets to the originally matched conditions, e.g.
large gradients, the larger defocusing forces leads to internal collimation, which in some instances leads to e.g. reduction in emittance. 
One should of course investigate and optimize a structure based on the desired final bunch characteristics and injection constraints; however the large operational
range of a single structure implies broad and stable operation for a single matching point. Some notable operating points include, 
($\sigma_z$, $\epsilon_r$, $\sigma_E$, $Q$)=(1.2~$\mu$m, 250~nm, 47.6~keV, 100~fC) for ($\phi$, $E_0$)=(79.3~deg, 106.875~MV/m).
The shortest bunch length acheived in our scans for the associated 
structure was ($\sigma_z$, $\epsilon_r$, $\sigma_E$, $Q$)=(730~nm, 158~nm, 83~keV, 80~fC) for the operational point ($\phi$, $E_0$)=(48.8~deg, 123.75~MV/m); smaller energy spreads
can be reached also, at the expense of other final parameters.

In summary, we have proposed adiabatically-tapered dielectric-lined waveguides to accelerate and manipulate low-energy charged particles with 
non-relativistic field strengths i.e. in the low-$\alpha$ regime. We hypothesized an ansatz for the transient field equations and support them with
Maxwell's equations and computer simulation.  We implemented the fields directly into {\sc astra} and perform beam dynamics simulations for a low-energy
electron bunch accelerating in a 1~mm field with 100~MV/m. The derived structure supports non-phase-slipping trajectories for various input powers; offsets in
the initial parameters leads to very similar final energies but with a wide variety of other bunch properties, notably small bunch lengths and energy spreads.
We presented a very simple beam matching scheme to accommodate the strong-defocusing forces in the early stages of acceleration.
Our derivation of Eq.~\ref{eq:limits} is independent of $\epsilon_r$ and is therefore a more general relation to the evolution of a traveling mode and
may appeal to e.g. plasma acceleration with smaller field gradients and potentially higher repetition rates via tapered density profiles.

The proposed setup could be further optimized given a specific application. The beam matching could be optimized via the positions and strengths between 
the elements to further minimize emittance growth for a given charge. Likewise, larger charges could be accelerated with higher injection energies, longer accelerating
wavelengths, and larger accelerating field strengths.  Acceleration with lower initial energies can be acheived with larger dielectric permittivities, and could possibly
realize a standalone relativistic electron source. A performance analysis given fabrication imperfections is also required and will especially be critical to the scaling 
of the concept to optical wavelengths, e.g. as needed for DLAs.
Finally, exploring alternative tapering profiles could open other applications of the proposed tapered accelerator including, e.g., the development of novel beam-manipulation techniques.

FL is thankful to DESY's accelerator division for support to continue this ongoing work. This project is funded by the European Union's Horizon 
2020 Research and Innovation programme under Grant Agreement No. 730871 and also supported by the European Research Council (ERC) under the European 
Union's Seventh Framework Programme (FP/2007-2013)/ERC  AXSIS Grant agreement No. 609920. PP is sponsored by US NSF grant  PHY-1535401.
We acknowledge the use of DESY's high-performance computing center {\sc Maxwell} and are grateful to F. Schluenzen for support. We thank M. Fakhari for providing 
the S-band gun design.

\bibliography{taper-accel}% Produces the bibliography via BibTeX.

\begin{thebibliography}{15}%
\makeatletter
\providecommand \@ifxundefined [1]{%
 \@ifx{#1\undefined}
}%
\providecommand \@ifnum [1]{%
 \ifnum #1\expandafter \@firstoftwo
 \else \expandafter \@secondoftwo
 \fi
}%
\providecommand \@ifx [1]{%
 \ifx #1\expandafter \@firstoftwo
 \else \expandafter \@secondoftwo
 \fi
}%
\providecommand \natexlab [1]{#1}%
\providecommand \enquote  [1]{``#1''}%
\providecommand \bibnamefont  [1]{#1}%
\providecommand \bibfnamefont [1]{#1}%
\providecommand \citenamefont [1]{#1}%
\providecommand \href@noop [0]{\@secondoftwo}%
\providecommand \href [0]{\begingroup \@sanitize@url \@href}%
\providecommand \@href[1]{\@@startlink{#1}\@@href}%
\providecommand \@@href[1]{\endgroup#1\@@endlink}%
\providecommand \@sanitize@url [0]{\catcode `\\12\catcode `\$12\catcode
  `\&12\catcode `\#12\catcode `\^12\catcode `\_12\catcode `\%12\relax}%
\providecommand \@@startlink[1]{}%
\providecommand \@@endlink[0]{}%
\providecommand \url  [0]{\begingroup\@sanitize@url \@url }%
\providecommand \@url [1]{\endgroup\@href {#1}{\urlprefix }}%
\providecommand \urlprefix  [0]{URL }%
\providecommand \Eprint [0]{\href }%
\providecommand \doibase [0]{http://dx.doi.org/}%
\providecommand \selectlanguage [0]{\@gobble}%
\providecommand \bibinfo  [0]{\@secondoftwo}%
\providecommand \bibfield  [0]{\@secondoftwo}%
\providecommand \translation [1]{[#1]}%
\providecommand \BibitemOpen [0]{}%
\providecommand \bibitemStop [0]{}%
\providecommand \bibitemNoStop [0]{.\EOS\space}%
\providecommand \EOS [0]{\spacefactor3000\relax}%
\providecommand \BibitemShut  [1]{\csname bibitem#1\endcsname}%
\let\auto@bib@innerbib\@empty
%</preamble>
\bibitem [{\citenamefont {Kim}(1989)}]{kwangje}%
  \BibitemOpen
  \bibfield  {author} {\bibinfo {author} {\bibfnamefont {K.-J.}\ \bibnamefont
  {Kim}},\ }\href {\doibase 10.1016/0168-9002(89)90688-8} {\bibfield  {journal}
  {\bibinfo  {journal} {Nucl. Instr. Meth. Phys. Res. A}\ }\textbf {\bibinfo
  {volume} {275}},\ \bibinfo {pages} {201 } (\bibinfo {year}
  {1989})}\BibitemShut {NoStop}%
\bibitem [{\citenamefont {Rosenzweig}\ and\ \citenamefont
  {Colby}(1995)}]{rosenswag}%
  \BibitemOpen
  \bibfield  {author} {\bibinfo {author} {\bibfnamefont {J.}~\bibnamefont
  {Rosenzweig}}\ and\ \bibinfo {author} {\bibfnamefont {E.}~\bibnamefont
  {Colby}},\ }\href {\doibase 10.1063/1.48260} {\bibfield  {journal} {\bibinfo
  {journal} {AIP Conference Proceedings}\ }\textbf {\bibinfo {volume} {335}},\
  \bibinfo {pages} {724} (\bibinfo {year} {1995})}\BibitemShut {NoStop}%
\bibitem [{\citenamefont {Piot}\ \emph {et~al.}(2003)\citenamefont {Piot},
  \citenamefont {Carr}, \citenamefont {Graves},\ and\ \citenamefont
  {Loos}}]{piot}%
  \BibitemOpen
  \bibfield  {author} {\bibinfo {author} {\bibfnamefont {P.}~\bibnamefont
  {Piot}}, \bibinfo {author} {\bibfnamefont {L.}~\bibnamefont {Carr}}, \bibinfo
  {author} {\bibfnamefont {W.~S.}\ \bibnamefont {Graves}}, \ and\ \bibinfo
  {author} {\bibfnamefont {H.}~\bibnamefont {Loos}},\ }\href {\doibase
  10.1103/PhysRevSTAB.6.033503} {\bibfield  {journal} {\bibinfo  {journal}
  {Phys. Rev. ST Accel. Beams}\ }\textbf {\bibinfo {volume} {6}},\ \bibinfo
  {pages} {033503} (\bibinfo {year} {2003})}\BibitemShut {NoStop}%
\bibitem [{\citenamefont {Floettmann}(2015)}]{klausrf}%
  \BibitemOpen
  \bibfield  {author} {\bibinfo {author} {\bibfnamefont {K.}~\bibnamefont
  {Floettmann}},\ }\href {\doibase 10.1103/PhysRevSTAB.18.064801} {\bibfield
  {journal} {\bibinfo  {journal} {Phys. Rev. ST Accel. Beams}\ }\textbf
  {\bibinfo {volume} {18}},\ \bibinfo {pages} {064801} (\bibinfo {year}
  {2015})}\BibitemShut {NoStop}%
\bibitem [{\citenamefont {Plettner}\ \emph {et~al.}(2006)\citenamefont
  {Plettner}, \citenamefont {Lu},\ and\ \citenamefont {Byer}}]{plettner}%
  \BibitemOpen
  \bibfield  {author} {\bibinfo {author} {\bibfnamefont {T.}~\bibnamefont
  {Plettner}}, \bibinfo {author} {\bibfnamefont {P.}~\bibnamefont {Lu}}, \ and\
  \bibinfo {author} {\bibfnamefont {R.~L.}\ \bibnamefont {Byer}},\ }\href
  {\doibase 10.1103/PhysRevSTAB.9.111301} {\bibfield  {journal} {\bibinfo
  {journal} {Phys. Rev. ST Accel. Beams}\ }\textbf {\bibinfo {volume} {9}},\
  \bibinfo {pages} {111301} (\bibinfo {year} {2006})}\BibitemShut {NoStop}%
\bibitem [{\citenamefont {Breuer}\ and\ \citenamefont
  {Hommelhoff}(2013)}]{breuer}%
  \BibitemOpen
  \bibfield  {author} {\bibinfo {author} {\bibfnamefont {J.}~\bibnamefont
  {Breuer}}\ and\ \bibinfo {author} {\bibfnamefont {P.}~\bibnamefont
  {Hommelhoff}},\ }\href {\doibase 10.1103/PhysRevLett.111.134803} {\bibfield
  {journal} {\bibinfo  {journal} {Phys. Rev. Lett.}\ }\textbf {\bibinfo
  {volume} {111}},\ \bibinfo {pages} {134803} (\bibinfo {year}
  {2013})}\BibitemShut {NoStop}%
\bibitem [{\citenamefont {Kapchinskii}\ and\ \citenamefont
  {Teplvakov}(1970)}]{rfq}%
  \BibitemOpen
  \bibfield  {author} {\bibinfo {author} {\bibfnamefont {I.~M.}\ \bibnamefont
  {Kapchinskii}}\ and\ \bibinfo {author} {\bibfnamefont {V.~A.}\ \bibnamefont
  {Teplvakov}},\ }\href
  {http://www.slac.stanford.edu/pubs/slactrans/trans01/slac-trans-0099.pdf}
  {\bibfield  {journal} {\bibinfo  {journal} {Prib. Tekh. Eksp.}\ }\textbf
  {\bibinfo {volume} {2}} (\bibinfo {year} {1970})}\BibitemShut {NoStop}%
\bibitem [{\citenamefont {Lemery}\ and\ \citenamefont
  {Piot}(2015)}]{lemeryIPAC1}%
  \BibitemOpen
  \bibfield  {author} {\bibinfo {author} {\bibfnamefont {F.}~\bibnamefont
  {Lemery}}\ and\ \bibinfo {author} {\bibfnamefont {P.}~\bibnamefont {Piot}},\
  }in\ \href
  {http://accelconf.web.cern.ch/AccelConf/IPAC2015/papers/wepwa070.pdf} {\emph
  {\bibinfo {booktitle} {Proc. 6th International Particle Accelerator
  Conference (IPAC'15), Richmond, VA, USA, 2015}}}\ (\bibinfo  {publisher}
  {JACoW},\ \bibinfo {address} {Geneva, Switzerland},\ \bibinfo {year} {2015})\
  pp.\ \bibinfo {pages} {2664--2666}\BibitemShut {NoStop}%
\bibitem [{\citenamefont {Lemery}\ \emph {et~al.}(2017)\citenamefont {Lemery},
  \citenamefont {Floettmann}, \citenamefont {Piot},\ and\ \citenamefont
  {Kaertner}}]{lemeryIPAC2}%
  \BibitemOpen
  \bibfield  {author} {\bibinfo {author} {\bibfnamefont {F.}~\bibnamefont
  {Lemery}}, \bibinfo {author} {\bibfnamefont {K.}~\bibnamefont {Floettmann}},
  \bibinfo {author} {\bibfnamefont {P.}~\bibnamefont {Piot}}, \ and\ \bibinfo
  {author} {\bibfnamefont {F.~X.}\ \bibnamefont {Kaertner}},\ }in\ \href
  {http://jacow.org/ipac2017/papers/wepab123.pdf} {\emph {\bibinfo {booktitle}
  {Proc. of International Particle Accelerator Conference (IPAC'17),
  Copenhagen, Denmark, 14-19 May, 2017}}},\ \bibinfo {series and number}
  {\bibinfo {number} {8}}\ (\bibinfo  {publisher} {JACoW},\ \bibinfo {address}
  {Geneva, Switzerland},\ \bibinfo {year} {2017})\ pp.\ \bibinfo {pages}
  {2861--2864}\BibitemShut {NoStop}%
\bibitem [{\citenamefont {Floettmann}(2011)}]{astra}%
  \BibitemOpen
  \bibfield  {author} {\bibinfo {author} {\bibfnamefont {K.}~\bibnamefont
  {Floettmann}},\ }\href {http://www.desy.de/~mpyflo/} {\enquote {\bibinfo
  {title} {Astra - a space charge tracking algorithm},}\ } (\bibinfo {year}
  {2011})\BibitemShut {NoStop}%
\bibitem [{\citenamefont {Kaertner}\ \emph {et~al.}(2016)\citenamefont
  {Kaertner} \emph {et~al.}}]{axsis}%
  \BibitemOpen
  \bibfield  {author} {\bibinfo {author} {\bibfnamefont {F.~X.}\ \bibnamefont
  {Kaertner}} \emph {et~al.},\ }\href {\doibase
  https://doi.org/10.1016/j.nima.2016.02.080} {\bibfield  {journal} {\bibinfo
  {journal} {Nucl. Instr. Meth. Phys. Res. A}\ }\textbf {\bibinfo {volume}
  {829}},\ \bibinfo {pages} {24 } (\bibinfo {year} {2016})},\ \bibinfo {note}
  {2nd European Advanced Accelerator Concepts Workshop - EAAC 2015}\BibitemShut
  {NoStop}%
\bibitem [{\citenamefont {Frankel}(1947)}]{Frankel1947}%
  \BibitemOpen
  \bibfield  {author} {\bibinfo {author} {\bibfnamefont {S.}~\bibnamefont
  {Frankel}},\ }\href {\doibase 10.1063/1.1697821} {\bibfield  {journal}
  {\bibinfo  {journal} {Journal of Applied Physics}\ }\textbf {\bibinfo
  {volume} {18}},\ \bibinfo {pages} {650} (\bibinfo {year} {1947})}\BibitemShut
  {NoStop}%
\bibitem [{\citenamefont {Daoud}\ \emph {et~al.}(2017)\citenamefont {Daoud},
  \citenamefont {Floettmann},\ and\ \citenamefont {Miller}}]{klausCompactGun}%
  \BibitemOpen
  \bibfield  {author} {\bibinfo {author} {\bibfnamefont {H.}~\bibnamefont
  {Daoud}}, \bibinfo {author} {\bibfnamefont {K.}~\bibnamefont {Floettmann}}, \
  and\ \bibinfo {author} {\bibfnamefont {D.}~\bibnamefont {Miller}},\ }\href
  {\doibase 10.1063/1.4979970} {\bibfield  {journal} {\bibinfo  {journal}
  {Structural Dynamics}\ }\textbf {\bibinfo {volume} {4}},\ \bibinfo {pages}
  {044016} (\bibinfo {year} {2017})}\BibitemShut {NoStop}%
\bibitem [{\citenamefont {Gehrke}(2013)}]{gehrke}%
  \BibitemOpen
  \bibfield  {author} {\bibinfo {author} {\bibfnamefont {T.}~\bibnamefont
  {Gehrke}},\ }\emph {\bibinfo {title} {{D}esign of {P}ermanent {M}agnetic
  {S}olenoids for {REGAE}}},\ \href {http://bib-pubdb1.desy.de/record/154592}
  {Master's thesis},\ \bibinfo  {school} {University of Hamburg} (\bibinfo
  {year} {2013})\BibitemShut {NoStop}%
\bibitem [{cst()}]{cst}%
  \BibitemOpen
  \href@noop {} {\enquote {\bibinfo {title} {Computer simulation technology
  (cst)},}\ }\BibitemShut {NoStop}%
\end{thebibliography}%

\end{document}